\documentstyle[11pt,psfig,aaspp4]{article}
\def\msol{\hbox{\kern 0.20em $M_\odot$}}
\def\lsol{\hbox{\kern 0.20em $L_\odot$}}
\def\rsol{\hbox{\kern 0.20em $R_\odot$}}
\def\sr{\hbox{\kern 0.20em sr}}
\def\srmu{\hbox{\kern 0.20em sr$^{-1}$}}

\def\g{\hbox{\kern 0.20em g}} 
\def\gmu{\hbox{\kern 0.20em g$^{-1}$}}
\def\kg{\hbox{\kern 0.20em kg}}
\def\pc{\hbox{\kern 0.20em pc}} 

\def\mum{\hbox{\kern 0.20em $\mu$m}}
\def\mumd{\hbox{\kern 0.20em $\mu$m$^{-2}$}}
\def\cm{\hbox{\kern 0.20em cm}}
\def\m{\hbox{\kern 0.20em m}} 
\def\km{\hbox{\kern 0.20em km}}
\def\nm{\hbox{\kern 0.20em nm}}

\def\s{\hbox{\kern 0.20em s}} 
\def\h{\hbox{\kern 0.20em h}}
\def\sec{\hbox{\kern 0.20em sec}}
\def\min{\hbox {\kern 0.20em min}}
\def\smu{\hbox{\kern 0.20em s$^{-1}$}}
\def\smd{\hbox{\kern 0.20em s$^{-2}$}}
\def\an{\hbox{\kern 0.20em an}} 
\def\anmu{\hbox{\kern 0.20em an$^{-1}$}}
\def\deg{\hbox{\kern 0.20em $^{\rm o}$}}
\def\yr{\hbox{\kern 0.20em yr}} 
\def\yrmu{\hbox{\kern 0.20em yr$^{-1}$}}
\def\Myr{\hbox{\kern 0.20em Myr}} 
\def\Mymu{\hbox{\kern 0.20em Myr$^{-1}$}}

\def\K{\hbox{\kern 0.20em K}}  

\def\pcmu{\hbox{\kern 0.20em pc$^{-1}$}}
\def\pcmd{\hbox{\kern 0.20em pc$^{-2}$}}
\def\pcmt{\hbox{\kern 0.20em pc$^{-3}$}}
\def\kms{\hbox{\kern 0.20em km\kern 0.20em s$^{-1}$}}
\def\kmpd{\hbox{\kern 0.20em km$^{2}$}}
\def\kpc{\hbox{\kern 0.20em kpc}}
\def\cms{\hbox{\kern 0.20em cm\kern 0.20em s$^{-1}$}}
\def\erg{\hbox{\kern 0.20em erg}}
\def\ergs{\hbox{\kern 0.20em erg}}
\def\cmpd{\hbox{\kern 0.20em cm$^2$}}
\def\cmmd{\hbox{\kern 0.20em cm$^{-2}$}}
\def\cmms{\hbox{\kern 0.20em cm$^{-6}$}}
\def\cmpt{\hbox{\kern 0.20em cm$^3$}}
\def\cmmt{\hbox{\kern 0.20em cm$^{-3}$}}
\def\mpd{\hbox{\kern 0.20em m$^2$}}
\def\mmd{\hbox{\kern 0.20em m$^{-2}$}}
\def\mpt{\hbox{\kern 0.20em m$^3$}}
\def\mmt{\hbox{\kern 0.20em m$^{-3}$}}
\def\mjy{\hbox{\kern 0.20em mJy}}
\def\Mj{\hbox{\kern 0.20em MJy}}
\def\jy{\hbox{\kern 0.20em Jy}}
\def\ghz{\hbox{\kern 0.20em GHz}}

\def\G{\hbox{\kern 0.20em G}}
\def\muG{\hbox{\kern 0.20em $\mu$G}}

\def\thco{\hbox{${}^{13}$CO}}

\def\ceio{\hbox{C${}^{18}$O}}

\def\htwo{\hbox{H${}_2$}}
\def\h13cop{\hbox{H$^{13}$CO$^{+}$}}
\def\hcop{\hbox{HCO$^{+}$}}

\newcommand{\jonetozero}{\hbox{$J=1\rightarrow 0$}}
\newcommand{\jtwotoone}{\hbox{$J=2\rightarrow 1$}} 
\newcommand{\jthreetotwo}{\hbox{$J=3\rightarrow 2$}}
\newcommand{\jfivetofour}{\hbox{$J=5\rightarrow 4$}}

\begin{document}

\newcommand{\jfourteen}{\hbox{$J=14\rightarrow 13$}}

\title{Pre-Orion Cores in the Trifid Nebula}

\author{Bertrand Lefloch\altaffilmark{1,2}
and Jos\'e Cernicharo\altaffilmark{1}}

\altaffiltext{1}
{Consejo Superior de Investigaciones Cient\'{\i}ficas,
Instituto de Estructura de la Materia, Serrano 123, E-28006 Madrid}
\altaffiltext{2}
{Laboratoire d'Astrophysique, Observatoire de Grenoble, BP 53,
F-38041, Grenoble Cedex 9, France}

\begin{abstract}
The Trifid nebula is a young H{\small II} region undergoing a burst 
of star formation. In this article, we report on far-infrared and millimeter
continuum and line observations of  several massive  and 
bright protostellar sources in the vicinity of the exciting 
star of the nebula, just behind the ionization front. These objects are 
probably young protostars (Class 0) and are associated with very massive 
cores ($M\sim 8-90 \msol$) powering young 
energetic outflows. Analysis of the far-infrared emission in the 
$45-200\mum$ range from ISO/LWS data shows that they are embedded in cold 
dense material. 
Inspection of their physical properties suggest that they are
similar to the dust protostellar cores observed in Orion, although at an 
earlier evolutionary ``Pre-Orion'' stage. The cores are embedded in a 
compressed layer of dense gas. Based on comparison with the models, we 
find that the cores could have formed from the fragmentation of the layer
and that the birth of the protostars was triggered by the expansion
of the Trifid nebula. 
\end{abstract}
 

\lefthead{Lefloch \& Cernicharo}
\righthead{Pre-Orion Cores in the Trifid Nebula}

\section{Introduction}

The presence of numerous protostellar objects in the bright-rimmed 
globules 
inside or at the border of H{\small II} regions has been widely reported. 
Based on the analysis of their IRAS colors, Sugitani et al. (1991)
concluded that theses sources could be preferentially forming 
intermediate-mass stars. They suggested that star formation triggered 
by the radiatively-driven implosion of bright-rimmed clouds could 
be a very efficient mechanism, contributing up to the formation of 
$\sim 5\%$ of the total stellar mass of the Galaxy. 
 
The few detailed studies of the molecular content 
and the continuum emission of protostellar cores in bright-rimmed globules
have reported the presence of intermediate-mass (AeBe) stars alone 
(Cernicharo et al. 1992) or accompanied by a cluster of low-mass stars 
(Megeath et al. 1996; Lefloch et al. 1997). 
However, hardly anything is known about the 
first evolutionary stages of these objects, mainly because  their  
search is based on IRAS color-color diagram. Hence we are led to 
deal with already evolved protostars which radiate a non-negligible part of 
their luminosity in the infrared (sources of type I/II in the 
classification of Lada, 1985) and have typical ages of a few $10^5\yr$ or 
more. Also, this characterization did not consider  the multiplicity of the
protostellar objects  in the infrared source.

However, as pointed out by Sugitani et al. (1991), bright-rimmed globules 
are found in relatively old H{\small II} regions, with ages of a few 
$\Myr$. These ionized regions are well-developed, with typical sizes 
$\geq 60'$, $18-36\pc$ at typical distances of $1-2\kpc$. 
The first drawback is that it is necessary to map/study huge areas in order 
to determine the whole protostellar content.
The second drawback is that bright-rimmed globules are typically found 
at rather large distances from the ionizing stars and experience a reduced 
UV field; hence,   
it is difficult to discriminate between a star formation 
induced by  the radiatively-driven evolution of the parent 
core and a globule which is spontaneously evolving and 
has already started to form stars when it is hit by the ionization front. 

With the aim of better understanding the impact of a strong UV field 
on star formation, and given the limitations of the previous studies, 
we have started a systematic multiwavelength study 
of a young H{\small II} region~: the Trifid (Cernicharo et al. 1998, 
hereafter CL98). It appears as a small dusty nebula of $10\arcmin$ diameter
immersed in a large CO molecular cloud (first reported by O'Dell, 1966) 
with a dynamical age of $0.3-0.4\Myr$ (see Sect.~7.2). 
Adopting a heliocentric distance of $1.68\kpc$ (Lynds et al. 1985),
the molecular cores on the border of the H{\small II} region 
lie at distances of $\sim 2\pc$ from the 
ionizing star. This is almost 5 to 10 times closer than for the 
sources studied by Sugitani (1991). The nebula is excited by HD~164492A, 
an O7V-type star of Ly-c luminosity $L_i= 7\times 10^{48}\smu$ (Panagia 1973). 

As can be seen from optical plates (Figure~1), the Southwest region of the 
nebula is
characterized by heavy obscuration and large amounts of dust.
The ionization front of the nebula impinges on a large
massive cloud ($M\geq 1300\msol$), first reported by Chaisson \& Willson 
(1975) in their study of the low-density gas content of the Trifid. 
They suggested 
that the cloud was undergoing a large-scale gravitational collapse and
could be a nest of star formation. More direct evidence 
for protostellar activity in the Trifid was recently shown by CL98, who 
detected an Herbig-Haro jet emerging from a cometary globule (dubbed TC2) 
in the 
Southern part of the  nebula (see also Rosado et al. 1999). They also 
reported the presence of two embedded sources (TC3 and TC4) in the 
Southwest molecular cloud. 

We present here a detailed study of the embedded sources reported 
by CL98 in the Southwest cloud, and show that they are similar to the 
protostellar cores  found in the  Orion nebula.

\section{Observations}

\subsection{The millimeter observations}

Observations of the 1.25mm continuum emission in the Trifid were carried out
in March 1996 and March 1997 at the IRAM 30m-telescope (Pico Veleta, Spain)
using the MPIfR 19-channel bolometer array. The beam size of the telescope 
is $11\arcsec$. The final map was obtained by combining several individual
fields, centered on the brightest condensations of the nebula. 
Each field was scanned by moving the telescope in the azimuth direction 
at a speed of $4\arcsec$ per sec and using a chopping secondary at 2Hz. 
The throw was chosen between 35\arcsec\ and 60\arcsec depending on the size
of the structures to be mapped. Subsequent azimuth scans were 
displaced by $4\arcsec$. With a sampling rate of 1/3 of a telescope beam, 
higher than the Nyquist frequency, we achieve an angular resolution of one half
telescope beam (6\arcsec). 
We adopted the coordinates of the central star as determined from
Hipparcos~: $\alpha_{2000}= 18^{\rm h} 02^{\rm m} 
23.55^{\rm s}$ $\delta_{2000}= -23^{\circ} 01\arcmin 51\arcsec$. 
The weather conditions were good and rather stable during the two
observing sessions. The final map yields an rms of
$8\mjy/11\arcsec$~beam. Hence, it is sensitive enough to detect at the
$3\sigma$ level condensations of $1\msol$ in one $11\arcsec$ telescope
beam ($0.09\pc$ at the distance of the Trifid). 

In July 1996 and 1997, we mapped the whole nebula and  the surrounding 
molecular gas with the IRAM 30m-telescope
over an area of 
$10\arcmin \times 10\arcmin$ at full sampling, using the On-The-Fly technique
in the millimeter lines of CO and $\thco$, and in the $\ceio\, \jonetozero$ 
transition. 
The technical details and the full maps will be presented and
discussed in a forthcoming paper (Lefloch et al. 2000).
Complementary observations with the 30m telescope were performed in the
$\jtwotoone$, \jthreetotwo\ and \jfivetofour\
lines of SiO and CS to trace the dense molecular gas in the direction of 
the main condensations detected in CO and in the continuum maps. 
All the lines were observed with a spatial sampling of $15\arcsec$ and
a velocity resolution of $0.2\kms$. 
We present here the data in the region of 
the Southwestern cloud (Figure~1). 
As the Trifid transits at low-elevation at Pico Veleta, pointing was checked 
every hour and corrections were always lower than $2\arcsec-3\arcsec$. 

\subsection{The ISO data}

The IR emission of the dust cores was observed with the instruments on  the
Infrared Space Observatory
ISO (see CL98 for details). We present here the data obtained with ISOCAM
(Cesarsky et al. 1996) at 3\arcsec\ resolution, taken with the LW10 filter 
($\lambda= 11.5\mum$, $\Delta\lambda= 7\mum$) and based on the most recent 
data reduction (pipeline 7). The images were corrected for field 
deformation. The noise in the region of 
the dust cores is $ \rm 1\mjy/3''$ pixel ($1\sigma$).
The ISOCAM image was positioned in right ascension and declination 
in two steps. In a first step, we took two optical images, one in 
$\rm H\alpha$ 
with the IAC80 telescope (Observatorio del Teide, Tenerife, Spain), and 
one in the [SII] $\lambda 6716-6730 \rm A$  lines with the Nordic Optical 
Telescope (NOT) at the 
Observatorio del Roque de los Muchachos (La Palma, Spain) (these images are
shown in CL98). 
Their astrometry was performed by identifying the sources reported on the
Tycho catalog. The positional errors are less than $1\arcsec$. 
In a second step, the optical sources with a counterpart in the 
$12\mum$ band were used to position the ISOCAM image. 
We found an agreement over the whole maps better than $3\arcsec$.

We have observed the dust continuum emission towards one dust core (TC4)
between 43 and $197\mum$ with the LWS spectrometer in grating mode. 
The spectral resolution of the LWS was 0.6$\mu m$ for $\lambda$ greater
than $90\mu m$ and 
0.3$\mu m$ below (R= 150/300). The size of the LWS beam (HPFW) is known to
vary from $\approx 60\arcsec$ to $70\arcsec$, depending on the detector
(E.~Caux, priv. comm.). In this work, we assume a gaussian beam of 
$70\arcsec$ HPFW. 
 
\section{The Continuum Emission}

\subsection{The dust millimeter cores}

We show in Fig.~1 a map of the 1.25mm continuum emission in the Trifid 
nebula, superposed on an $\rm H\alpha$ image taken with the IAC80 telescope 
(see CL98 for details), and the location of the 
Southwestern cloud where the embedded sources were discovered.
A magnified view, displayed on Fig.~2, shows the dust thermal emission
of two strongly peaked condensations (dubbed TC3 and TC4) at offset
positions (-246\arcsec, -221\arcsec) and (-150\arcsec,-246\arcsec)
respectively. To the Southeast of TC4 we detect two faint and smaller
condensations (dubbed TC4b and TC4c) at offset positions
(-126\arcsec, -300\arcsec) and (-90\arcsec, -320\arcsec).
The TC4-TC4c condensations  lay in a dust lane of $\approx 40\arcsec$ 
width (0.33\pc) and $160\arcsec$ length (1.3\pc), at a projected distance 
of $2.7\pc$ from the exciting star of the nebula. The dust lane is
essentially adjacent to the ionized gas region, as traced by the $\rm H\alpha$ 
emission, and constitutes 
one of the several fragments of a shell which bounds the H{\small II} region.
We note however a small overlap between the dust lane and the ionized
region. The geometry of the cores with respect to the nebula is analyzed 
in more detail in Sect.~4. 

The sources have sizes (estimated from the $50\%$ contour in the
millimeter flux map and after beam deconvolution) of 0.11\pc\ and 0.16\pc\
for TC4c and TC4b respectively, and 0.2\pc\ for TC4 and TC3, with
a smooth flux distribution down to 6\arcsec\ (0.05\pc\ at the distance of
the Trifid). They are well resolved by the
11\arcsec\ beam of the 30m telescope though only marginally in the case of
TC4b-TC4c. The peak flux of the dust cores is 0.39~Jy/11\arcsec\ beam and
0.29~Jy/11\arcsec\ beam for TC3 and TC4 respectively.
Based on the analysis of the far-infrared continuum emission of TC4,
we estimate
a dust temperature of $20\K$ and a dust spectral index $\beta= 2$ (see below).
Since our millimeter mapping shows a large similarity between the TC3 and
TC4, we also applied these dust parameters to derive the properties of
TC3.
Adopting a  dust opacity $\kappa_{250}= 0.1\cmpd\gmu$ at $250\mu m$,
and integrating over the $50\%$ contour level, we derive core masses 
ranging from  $8\msol$ (for TC4c) to $90\msol$ (for TC3). The hydrogen 
column densities estimated
at the flux peak of the cores range from 0.6 to
$4\times 10^{23}\cmmd$  (see Table~\ref{core}). Assuming a ``normal''
extinction ratio this corresponds to visual extinctions $\rm A_v$ between
34 (TC4c) and 192 (TC3). If we adopt the ``anomalous'' extinction ratio
$\rm A_v/E(B-V)= 5.1$ (derived by Lynds, Canzian \& O'Neil (1985) from 
$\rm H\alpha$ and $\rm H\beta$ measurements of the Trifid) we obtain 
even higher visual extinctions, in the range 60 -- 325. 
The physical properties of the cores are summarized in Table~\ref{core}. 

TC3 appears to be immersed in an extended envelope of
low-emissivity. This envelope is elongated in the North-South direction, 
and has a typical size of $1.5\times 1\pc$ ($3'\times 2'$), 
defined from the contour level at $15\mjy/15\arcsec \, \rm beam$
in the degraded map (Fig.~2). The flux distribution appears somewhat 
irregular (``clumpy''), exhibiting two other small condensations
close to TC3. 
Assuming the same dust physical properties for the envelope as for TC3 
we derive a total mass of $750\msol$.

\subsection{The mid to far-infrared emission}

The low angular resolution  and sensitivity of
IRAS provided only poorly contrasted images of the Trifid. Subsequently,
it also failed in detecting point sources in the nebula. 
We show in Figure~3 a view of the TC3-TC4 region observed with ISOCAM
at $12\mum$ (LW10 filter) and a pixel size of $3\arcsec$. 
We detect a diffuse extended component 
of flux ranging between 70 and $150\Mj\srmu$. The rms of the map 
in the region is $5\Mj\srmu$, or $1\mjy$ in a $3\arcsec$ pixel. The region
of minimum emissivity coincides with the dense molecular gas layers
at the border of the nebula. We also find evidence for a few point-like
sources in the surrounding of TC3 and TC4. These IR sources probably 
correspond to newly born stars, formed in the large
burst which accompanied the formation of the nebula. 

In the Southern part of the TC3 envelope, two objects are apparent 
at offset positions (-257\arcsec,-326\arcsec) and
(-248\arcsec,-300\arcsec). Close to TC4, we discovered two sources, unnoticed
previously~: the first one lies at the offset 
position (-152\arcsec,-205\arcsec) and the second one lies $\sim 6\arcsec$ 
North of the TC4 dust peak, at the offset position (-148\arcsec,-237\arcsec). 
The latter has a peak flux of $170\Mj\srmu$ and an integrated flux of
$0.25\jy$. It is marginally resolved 
by ISOCAM~: we estimate a size of $3\arcsec$ (0.024\pc), comparable to the 
angular resolution of the observations. 
The distance between the IR point source and the TC4 dust peak 
 is comparable to the relative position uncertainties of the
$1300\mu m$ and $12\mu m$ maps, hence we cannot exclude that, instead of 
originating from another  object on a closeby line-of-sight, 
this emission could be associated with the TC4 dust peak.
In agreement with CL98, we find only a smooth flux distribution and 
no point-like emission towards TC3. 

The far-infrared emission towards TC4 was observed with the LWS
spectrometer in the range $45-197\mu m$, in the course of a one-dimensional
raster across the nebula (not shown here). 
The emission line spectrum of TC4 (Fig.~4) is dominated by
atomic and ionic lines (O{\small I}, O{\small III}, N{\small II}, 
N{\small III}, C{\small II}), excited mainly
in the Photon-Dominated Region (PDR) and the ionized gas of the 
nebula (Fig.~4). In particular,
even the CO rotational transition of lowest energy
accessible at $185.9\mu m$ remains undetected. 
This stresses the fact that the line emission originates from a 
cold molecular component.  We used the LWS data to build the spectral 
energy distribution (SED) 
of TC4, in order to estimate the dust physical conditions in the region. 
Three terms contribute to the flux collected in the direction of TC4~:
the TC4 core, the PDR and the H{\small II} region. 

We first derive a lower
limit to the SED by estimating and subtracting the contributions of the 
extended components, i.e. the PDR and the H{\small II} region to the flux
collected towards the TC4 core. 
These two contributions were estimated  from observing a reference position
located on the eastern side of the nebula~: 
$\alpha(2000)= 18^{h}02^{m}35.1^{s}$,
$\delta(2000)= -23^{\circ}03'52''$ (J2000). This position was chosen
because all the maps of various molecular tracers 
indicated only very weak or no emission at all in this direction. 
The raster shows only weak
variation of the emission in the range $45-90\mu m$ over the nebula, 
irrespective of
the presence of molecular gas on the line of sight. This suggests that the bulk
of the emission between 45 and $90\mu m$ rather originates from 
the H{\small II} region and the PDR. Hence, it is unlikely  that 
the contribution of the H{\small II} region and the PDR are much 
overestimated and it gives us confidence in the procedure followed. 
We cannot exclude however that a small mid-infrared contribution 
originating from some warm material  within the TC4 core itself might have
been missed through this procedure. 

After subtracting the contribution of the PDR and the H{\small II} region, 
the overall emission from TC4 can be satisfactorily 
fitted by a 
modified black-body with an opacity law $\tau_{\nu}\propto \nu^2$, for a 
hydrogen column density $\sim 2.1\,10^{23}\cmmd$ and a dust temperature
$T= 20\K$. Figure~4 shows that the fit is well constrained by the data. 
Some deviation is observed at $\lambda < 60\mu m$.
This excess probably arises from the contribution of the photon-dominated
region and the ISOCAM sources. 
Interestingly, the distribution peak appears to lay longwards of $200\mum$;
more measurements, especially in the submillimeter regime would help to  
better determine the dust properties in the far-infrared
range. Integrating under the fit to the spectral energy distribution (SED), 
we derive a bolometric luminosity of $520\lsol$ for TC4. This value is
probably a lower limit as discussed above. 

An upper limit to the bolometric luminosity can be obtained by fitting the
emission of the full unsubtracted LWS spectrum 
The emission between 
45 and $197\mu m$ can be reproduced quite closely by assuming the LWS beam 
to be filled by two components, modeled as a modified black-body~: 
a cold component of column density $\rm N_H= 1.3\times 10^{23}\cmmd$ at 20~K
with a spectral dust index $\beta= 2$, and a warmer component 
of lower column density $N_H= 2.5\times 10^{21}\cmmd$, at 40~K, 
with  $\beta= 1.2$, typical of the dust emission in the mid-IR 
(Hildebrand 1983).
Integrating under the fit (Fig~4), we obtained a total luminosity 
$L\simeq 2400\lsol$. As a conclusion, the bolometric luminosity of TC4 lies in
the range $520-2400\lsol$. 

As the bolometer bandpass is $\approx 70$~GHz, and adopting a bolometric
luminosity $L_{bol}= 520$, we derive a 
ratio $L_{bol}/(10^3\times L_{1.25})\sim 9$; such low values 
are typical of  young protostars (see e.g. Andr\'e et al. 1993) and indicate
that a non-negligible fraction of the mass is still to be accreted from the
envelope. The nature of TC3 and TC4 is further discussed in Sect.~5. 

\section{The molecular emission}

We first examine the overall geometry of the dust cores TC3-TC4 with
respect to the H{\small II} region, by analyzing the kinematics of  
the molecular material and its distribution with respect to the ionized 
gas. 

Figures 1 and 2 show that the dust lane
containing the dense cores TC3-TC4 {\em partly} overlaps with the ionized 
gas region. The high visual extinction of the  cores and the dust lane implies
that they are located behind the ionized region, on the rear side of the
nebula. 
The Southern part of the TC3 envelope appears free of any $\rm H\alpha$ 
emission, hence must lay below the border of the H{\small II} region. 

This is confirmed by the analysis of the molecular gas kinematics. 
The mapping of the molecular emission shows very complex profiles in
the low-density gas, as traced by CO (Fig.~5; Lefloch et al. 2000). 
The spectra display numerous kinematic components in the range 
$[-50;+50]\,\kms$, most of them of low column density. From a comparison 
with an optical image of the nebula (Fig.~1), we found that some of this 
material is seen in absorption against the H{\small II} region as
the well-known ``arms'' of the Trifid. 
The material on the front side has a  velocity $v_{lsr}$ 
in the range $0-10\kms$. 
There is hardly any molecular emission between 10 and $15\kms$. At
higher velocities, between 15 and $18\kms$, one finds again large column 
densities of material, over the area of the optical nebula. 
The material on the rear side emits at velocities 
larger than $15\kms$, in agreement with an expanding
motion for the H{\small II} region. 

The molecular emission of the TC3-TC4 cores peaks at velocities 
$v_{lsr}\sim 20-23\kms$, somewhat larger than the mean velocity of the 
rear side of the  nebula. Some molecular transitions, like the  
CO \jtwotoone\ line (see Fig.~5), are absorbed 
between 5 and $20\kms$. This velocity range implies that the absorbing 
material is located on the front side of the nebula. 

The low-density gas was already studied 
by Chaisson \& Willson (1975) in some lines of H$_2$CO 
(angular resolution of 6.6') and OH (angular resolution of 18'). 
They found a maximum of emission coinciding with the Southwestern 
molecular cloud. They reported a broadening of the lines and a velocity 
gradient towards the cloud center, where TC3 and TC4 are found, which 
they suggested as evidence of gravitational collapse. 

\subsection{The CS observations}

The region of TC3 and TC4 was mapped in the millimeter lines of CS. 
We show in Fig.~6 a map of the main-beam brightness temperatures 
detected in the $\jthreetotwo$ line. We found two centrally peaked 
condensations 
centered on the continuum flux peaks of the TC3 and TC4 cores.  
The lines are relatively bright with typical intensities of 1-2~K in all 
three transitions. Towards the center, the line profiles exhibit broad 
wings which are the signature of bipolar outflows whose properties are
studied in Sect.~5. The linewidths of the three transitions range 
between $2\kms$ (TC3) and $4\kms$ (TC4) and are larger than 
the lines observed in the low-density gas traced by \ceio\, 
 ($\sim 1.6\kms$), especially in the direction of the TC4. 
However, as discussed in Sect.~5, this apparent broadening could be 
due to the geometry of the outflows powered by the protostars. 

In TC4, the emission peaks at $\approx 22.3\kms$ (Fig.~5). 
The half-power contour 
delineates a round condensation of $40\arcsec$ diameter ($0.33\pc$). The 
$\jfivetofour$ line was detected at a few positions towards the center of the 
dust core. We derived the temperature, $\htwo$ volume density and CS column
density in the condensation peak from a Large Velocity Gradient 
(LVG) analysis of the three lines, 
using the collisional rates of Green \& Chapman (1978). 
In the central region, the lower $\jtwotoone$ and $\jthreetotwo$ transitions 
are mainly sensitive to the gas column density; hence these two lines  
do not constrain very severely the density and the temperature. We found  
as ``best solution'' a kinetic temperature of 20~K (in good agreement with the 
determination from the SED), an $\htwo$ volume density 
$\sim 6\times 10^5\cmmt$ and a column density 
$\rm N(CS)= 3.3\times 10^{13}\cmmd$. In the outer parts, the \htwo\ 
density and 
the CS column density were determined from the $\jtwotoone$ and $\jthreetotwo$ 
lines. We find little variations in the values derived across the core. 
Integrating over the HPFW contour and assuming 
an abundance $[\rm CS]= 10^{-9}$, we obtain a total mass 
$\rm M= 58\msol$, a value in very good agreement with the estimate from the
dust millimeter observations. 

In TC3, the \jthreetotwo\ emission reveals a flattened condensation
of $20\arcsec\times 14\arcsec$ ($0.16\times 0.11\pc$, after 
beam-deconvolution). Unlike TC4, the emission of the $\jfivetofour$ line 
is unresolved and coincides with the peak of 1.25mm continuum core, where we
detect some bright emission ($\rm T_{mb}= 1.8\K$); only some 
weak emission is detected elsewhere. The linewidths (HPFW) are $\sim
2\kms$, narrower than in the TC4 condensation 
($\Delta v\simeq 4\kms$). The \jthreetotwo\ line profiles are double-peaked 
in the center and East of TC3 (Fig.~5). One of the peaks is centered at 
the velocity 
of the TC4 core ($22.7\kms$) whereas the other one is centered at 
$20.7\kms$ (Fig.~5). The first component is spatially distributed between 
TC4 and TC3. In the higher \jfivetofour\ transition, the emission is 
centered at $20.7\kms$, similar to what is observed in 
other molecular tracers like SiO, which are more indicative of protostellar
activity. This shows that the \jthreetotwo\ line is not self-absorbed 
but that we detect a second component. Based on its velocity and its
spatial distribution, the latter appears to be nothing else but the 
gas layer containing TC4 superposed on the red wing of the TC3 outflow, 
which produces the bright emission observed. We find for 
this layer typical densities of $\sim 1-2\times 10^5\cmmt$ and column 
densities N(CS)$\sim 10^{13}\cmmt$.
Towards the TC3 core, we derive 
physical properties rather similar to those
reigning in TC4. At the brightness peak, we find a kinetic temperature 
of 20~K, 
a density $\simeq 1.6\times 10^6\cmmt$ and a column density $2\times
10^{13}\cmmt$. 
In the envelope, we find densities $\simeq 4-5\times 10^5\cmmt$ out to 
$30\arcsec$ North of the TC3 peak. However, only $15\arcsec$ South of TC3, 
the density has already decreased to $1.4\times 10^5\cmmt$. 
Similarly, we note a decrease from $2\times 10^{13}\cmmd$ to 
$\simeq 8\times 10^{12}\cmmd$ in the  column densities as one moves away 
from the TC3 peak.
Integrating the column densities over the HPFW contour and adopting a CS
abundance $\rm [CS]= 10^{-9}$, we derive a mass $M\simeq 12 \msol$. 

The above mass determinations are affected by the uncertainties on the
CS abundance in the Trifid nebula. The comparison with the results from 
the 1.25mm continuum observations is very satisfying in the case of TC4, 
but there is a noticeable discrepancy between both estimates in the case of 
TC3. A simple calculation of the core virial mass 
based on the extent of the CS emission and the non-thermal \ceio\ 
linewidth yields a value of $35\msol$ and $88\msol$ for TC3 and TC4
respectively. The three mass estimates of the TC4 core (millimeter
continuum, CS, virial) agree quite well. The situation is more contrasted
in the case of TC3. 

The kinetic support against gravity was not estimated from the CS
linewidths for the reasons explained above. Instead, it was calculated 
from the $\ceio\, \jonetozero$ line. Thanks to its low opacity 
and its lower critical density (see Sect.~4.2), the $\ceio\, \jonetozero$ 
transition probes larger amounts of ambient material in and around the 
cores and their envelope, without being too much 
sensitive to the outflowing gas. This is illustrated by the line 
profiles displayed on Fig.~5~: unlike CO, CS and SiO, the \ceio\ line does not
display any strong high-velocity wing . Hence, this linewidth is a better 
tracer of the non-thermal motions in the regions considered here. 

In order to estimate the physical conditions in the molecular gas
surrounding the TC3-TC4 condensations, we also observed the position
$(-240\arcsec,-120\arcsec)$, inside the gas filament on which 
the Trifid impinges and detected in the \ceio\ line  
(see below). We found a weak \jtwotoone\ line of $\sim 0.4$~K, and set an  
upper limit of 0.15~K to the $\jthreetotwo$ line. From the line ratio, we
derive an upper limit to the \htwo\ density of $3\times 10^4\cmmt$, 
and an average CS column density $\sim 10^{13}\cmmd$.  
  
\subsection{The \ceio\ observations}

The picture drawn by the \ceio\ emission is very different from the
high-density gas tracers like \hcop\ or CS. The bulk of the emission 
is comprised in the velocity range $17-23\kms$. The lines are bright
($\rm T_{mb}$ of a few K) and rather broad ($\Delta v\simeq 1.6-1.8\kms$). 
Figure~7 shows
the $\ceio\, \jonetozero$ emission integrated in channels of $0.5\kms$ 
between 17.5 and $23\kms$. 

In the velocity interval where the dense gas was observed -as traced by 
the CS lines-  
($20-22\kms$), there is no clear large-scale condensation centered on 
the TC3 and TC4 dust peaks.
We detect  a local maximum centered on TC4 and a condensation where 
TC3 peaks at its apex.
We note however a good agreement between the 
\ceio\ emission and the 1.25mm continuum map when the resolution of the
latter is degraded to $22\arcsec$.
Also, the gas column densities derived under the LTE hypothesis, assuming
an excitation temperature of 20~K, a standard abundance of
$2.5\times 10^{-7}$ (Cernicharo \& Gu\'elin 1987) and optically thin lines,
agree within a factor of 2 with the estimates from the dust emission,
except in the central region of the cores.
We find that the column densities derived from the line measurements tend to be
lower than those derived from the dust emission. The 
discrepancy is more and more pronounced in the densest layers of the core,
up to a factor of 5 at the peak of TC3. 
At this position, we estimate a ``degraded''  continuum flux of $0.24\jy$ in
one $22\arcsec$ beam, which corresponds to a column density
$\rm N(\ceio)= 7.5\times10^{15}\cmmd$.
Adopting the same density and temperature as for the dense cores, an LVG
calculation gives an opacity $\tau^{10}= 0.17$ for the \jonetozero\
transition. The low opacity of the $\ceio\, \jonetozero$ line suggests
that the discrepancy could be the signature of molecular depletion onto the 
dust grains.
We note  that in the protostellar cores of other star-forming regions
(e.g. IRAS4A and IRAS2 in NGC1333, Lefloch et al. 1998)
where evidences of molecular depletion have been reported, larger discrepancies
are found. Observations at higher  angular resolution could help clarify 
this point. 

On larger scales, the mapping of the region in the \ceio\ and CO lines 
shows that the gas emission 
is distributed in a very long filament on the rear side of the nebula, 
which stretches far away from it. This emission arises at lower 
velocities than those of the protostellar cores and can be seen 
on Figure~7 in the range $17.5-20.0\kms$. 
The morphology of the filament is very conspicuous in that it seems to
change in direction and bend around the TC3-TC4 cores, leaving a ``hole''
at their location. 
It is actually the whole layer 
detected in the dust millimeter emission and housing the TC4-TC4c
condensations which is unambiguously separated from the filament. 
The border of the filament around the cores is characterized by stronger 
gradients in the brightness map, especially at the Northwest of the TC3 
condensation. 
We estimated the \htwo\ column densities in the filament with the
assumption of optically thin lines in the LTE regime. We find low values 
of the order of $1-5\times 10^{21}\cmmd$.

\section{Bipolar Outflows around and TC3 and TC4}

We mapped the dust core region in the SiO \jtwotoone\ line 
as this molecule is a good tracer 
of the outflowing gas of young  and/or massive protostellar sources. 
We show in Fig.~\ref{fig8} a map of the SiO \jtwotoone\ 
velocity-integrated emission. We detected in the three millimeter SiO lines
two  compact bipolar outflows centered on
the TC3 and TC4 dust peaks at the velocities $v_{lsr}= 21.0\kms$ and 
$v_{lsr}= 23.5\kms$ respectively. However, due to lack of time, only a few
positions could be observed in the higher \jthreetotwo\ and \jfivetofour\ 
transitions. 

We found that the wings cover a total velocity range 
of 70\kms\ (from -10 to $+60\kms$) in both sources (see Fig.~5). 
Towards TC3, the wings have a size (beam deconvolved) of 35\arcsec\
(0.28\pc), and their centroids are separated by 15\arcsec, 
nearly one telescope beam at 2mm. The line profiles are typical 
of young high-velocity outflows~: a narrow line core with some low 
brightness emission at high-velocity (see left panel of Fig.~5). 
Towards TC4, the wings have a similar size but fully overlap 
(Fig.~\ref{fig8}), indicating that the outflow is oriented close 
to the line of sight. 
Such a geometry implies that the column densities of accelerated material
along the line of sight are larger; adding their contribution to the 
ambient material results in broader and more complex profiles. 
Indeed, the line profiles are broader in TC4 than in TC3~:
$2\kms$ versus $4\kms$ for  the CS \jthreetotwo\ line. This effect can be seen 
also on the SiO \jfivetofour\ profile on Fig.~5. 
In any case, given the compactness of the TC3 and TC4 outflows, 
they are probably very young. From the extent of the wings we estimate a
kinematical age of $6.8\times 10^3\yr$, a value typical for Class 0
sources. 

We first estimated the physical conditions in the gas from an LVG analysis
of the three transitions. The CS collision rates were adopted for  the 
SiO molecule. We used a velocity width of 4\kms\, corresponding to the core
of the lines. We attempted to determine the density, column 
density and kinetic temperature at the central position of TC3 and TC4. 
The temperature and density determinations are somewhat uncertain as
the \jtwotoone\ and \jthreetotwo\ are mostly sensitive to 
the column density. We found that the kinetic temperature which best allows 
to reproduce the brightness temperatures is definitely above 40~K and 
close to 60~K. We adopted this value as the gas kinetic temperature in the
derivation of the outflows parameters. 
For kinetic temperatures $T_k$ in the range 20-80~K, we find 
that the density, as determined from the $\jfivetofour$ and $\jthreetotwo$
lines, is comprised between $3\times 10^5\cmmt$ and $1.0\times 10^6\cmmt$, 
and the column density $\rm N(SiO)\simeq 5\times 10^{13}\cmmd$. In this regime,
the opacity $\tau^{21}$ varies in the range 0.02-0.05, so that 
the \jtwotoone\ line is optically thin. The excitation temperature of the 
line varies between 
$T_{ex}^{21}= 30$~K for a kinetic temperature $T_k= 40~K$, to 
$T_{ex}^{21}= 43$~K for $T_k= 80$~K; the line is thermalized only at low
temperatures. 
The same conclusions hold for TC4, where we find a slightly lower 
column density N(SiO)$\simeq 2.7\times 10^{13}\cmmd$. 

Towards the central position, we estimated the gas density in the wings 
from the ratio of the $\jthreetotwo / \jtwotoone$ brightness temperatures. 
The results are consistent
with an analysis done on the $\jfivetofour / \jthreetotwo$ ratio but it offers
the advantage of a higher signal-to-noise ratio. We find that the line 
ratio does not 
depend very much on the velocity; it varies in the range 1.1-1.4 for TC3
and 1.2-1.4 for TC4. This corresponds to densities in the  range 
$2.0-20\times 10^5\cmmt$ and $2.5-7.5\times 10^5\cmmt$, respectively,
for a kinetic temperature in the interval of 20-80~K. These densities
are of the same order as in the surrounding molecular gas, as determined
from the CS lines. 

Since the excitation temperature of \jtwotoone\ line does not vary too much
in the range of kinetic temperatures considered, and the line is optically
thin, we adopted an average value $T_{ex}^{21}= 40~K$  in order to derive 
the large-scale properties of the outflows in the LTE regime. 
We did not correct for the inclination of the outflows with respect to the
plane of the sky. 
As discussed above, the adopted excitation temperature is probably accurate 
only within 50\%. However, we believe it is well representative of the
physical conditions in the outflowing gas. 
Hence, we derived  the SiO abundance in the high-velocity gas from 
comparison of the CO and SiO column densities in the red
wing of TC3. As can be seen on Fig.~5, the CO red wing appears to be 
free from contamination by the ambient molecular cloud for $v\geq 24\kms$. 
At the typical densities encountered in the outflows, the CO millimeter 
lines are thermalized and we adopted an excitation temperature of 60~K 
for the \jtwotoone\ line. We also assumed a standard CO 
abundance of $10^{-4}$.
We derived $\rm [SiO]\simeq 2\times 10^{-8}$, a value rather high and 
typical of 
very young high-velocity flows (Lefloch et al. 1998), that we used 
to estimate the outflows parameters. We find total masses 
of 1.1 and $1.5\msol$, mechanical energies of $\sim 10^{45}\erg$ and  
luminosities of 1.2 and $1.1\lsol$ for the TC3 and TC4 outflows respectively. 
The physical parameters are summarized in
Table~\ref{outflow}. Both outflows appear very similar under the reasonable
assumptions that their SiO abundance and temperature are not too
different.

\section{TC3 and TC4 as Pre-Orion Cores}

\subsection{Temperature of the sources}

Our continuum observations show the presence of an extended cold 
envelope around protostars TC3 and TC4. However, we cannot 
exclude the presence of a warm component unresolved 
by the beam of the IRAM 30m telescope, since it gives only access to spatial 
structures larger than $0.05\pc$ at the distance of the Trifid (1.68\kpc).
Actually, observations of high-mass protostars in Orion reveal 
the presence of a hot core in these sources, a dense component with 
a size of $\sim 0.02\pc$ and a temperature of 70-100~K (see Table~2). 

Such component would remain unresolved in our continuum millimeter
observations; moreover, its contribution could have been ignored when we 
derived the far-infrared emission of TC4 from the LWS spectrum.
On the contrary, the angular resolution and the sensitivity of our  
$12\mu$m ISOCAM map are high enough to constrain the temperature 
of this warm component. Our $12\mu$m map shows one source with a 
size of $\sim 0.025\pc$ which could coincide with the TC4 dust peak taking 
into account the positional uncertainties.

We now assess if the ISOCAM fluxes measured towards the region are
compatible with the presence of a warm core embedded in a cold gas envelope
of column density similar to the peak value derived from the 
1.25mm continuum observations. The column density of the cold component 
could be higher than the estimates at the 1.25mm flux peak 
but it seems unlikely that it be {\em much} higher. We also
assumed that the warm source lies approximately at the center of the 
envelope. The absorption by the cold component was estimated from the 
visual extinctions derived towards TC3-TC4 (see Sect.~3.1), assuming an   
average extinction ratio $<A_{\lambda}/A_{v}>= 0.03$ for the LW10 band, 
based on the values tabulated by Savage \& Mathis (1979). 

Applying this procedure to the TC4 region, it comes out that an optically 
thick black-body 
of $2.5\arcsec$ diameter (0.02\pc) at a temperature $\rm T_d= 80~K$ 
provides a flux $\approx 0.1\jy$ at 
12$\mu$m, compatible with the $0.25\jy$ source detected 
close to the millimeter dust peak. 
In the case of TC3, there is no indication of a point-like source down to 
$1\mjy$ per $3\arcsec$ pixel. This sets an upper limit of $\sim 70\K$ for 
the warm (optically thick) component. 

These simple estimates show that the presence of a hot core in TC4 
is likely if the physical association with the $12\mu m$ is
real. The situation is less clear for TC3, and 
complementary observations in various molecular transitions  could help
determine more accurately the temperature towards the peak of the
condensation. Confirming the absence of a gas component significantly 
warmer than the TC3 envelope would mean that the central source is 
a true massive protostar.
It is interesting however that both TC3 and TC4 are still deeply
embedded in large amounts of cold dense material. Together with the 
young kinematical age of the outflows, this is suggestive of an early 
evolutionary age.

\subsection{Comparison with other young protostellar sources}

The Jeans length is $\sim 0.06\pc$ in the cores of TC3 and TC4~:
the presence of subcomponents in both cores is therefore 
possible in principle. However, the absence of apparent substructure down to 
$0.05\pc$ (one half telescope beam at 1.25mm) makes the 
presence of more than a few unresolved components in our 
observations very unlikely. This does not alter our  conclusions 
about the high mass of the protostellar cores in the Trifid. 

Comparison with other protostellar sources is somewhat hampered by the 
relatively large distance to the Trifid, as compared to the star-forming
regions where the youngest protostars have been found until now. 
The difference in size between TC3-TC4 and other protostellar sources 
(typically a factor 3 to 5) comes from the fact that we take into 
account the contribution of the extended envelope. For instance, the Cep\,E 
protostellar condensation has a size of $0.2\times 0.12\pc$ 
(Lefloch et al. 1996). 
TC3 and TC4 are more massive and brighter than most of the Class~0 
sources known to date (Bachiller 1996).  Their masses are  
more than one order of magnitude larger than the typical 
sources VLA1623, L1448-mm (see Table~1). They appear also more massive than
most of the protostars discovered by Chini et al. (1997) in Orion towards 
the OMC-2 and OMC-3 cores. 

The luminosity of TC4 compares 
well with the young intermediate-mass protostar NGC~7129-FIRS2 
(Eiroa et al. 1998). 
However the more massive envelope of TC3-TC4 and their lower mean core 
temperature ($20\K$ versus $35\K$ for NGC7129)
suggests an earlier evolutionary age. 
On the contrary, the similarity of the physical properties 
between the outflows of TC3-TC4 and  the sources studied by Shepherd \& 
Churchwell (1996), as well as the bolometric 
luminosity of TC4 (comparable to a $5\msol$ star), strongly suggests
that the Trifid protostars could be intermediate-mass objects. 

It is interesting to compare the line intensities of some high-density 
tracers like CS of SiO with what is observed in the massive cores of Orion.
Neglecting the structure of the emitting regions, a rough estimate 
for the expected CS and SiO lines
is provided by directly scaling the peak main-beam temperatures observed 
towards TC3 and TC4 at the distance of Orion ($500\pc$). 
As a result, the CS lines observed in TC3 would appear similar to those 
measured towards the massive cores detected of Mundy et al. (1986) in the 
OMC1 region (see Table~\ref{compar}). Similarly, the millimeter SiO lines 
would be $\sim 6-14\K$ whereas Martin-Pintado (1998) has detected lines of 
$17-23\K$ in the IRc2 core. 

Therefore, the central protostars in TC3 and TC4 look similar 
to the Orion protostellar cores. 
On the other hand, the Trifid cores are still deeply embedded in 
some material much colder than the Orion nebula. Presumably, the 
parental medium has not yet been warmed up neither by the UV radiation 
of the exciting star, nor by the central sources. 
The TC3 and TC4 cores appear to be in a ``pre-Orion'' stage, 
characteristic of the early evolution of a nascent H{\small II} region~:
the massive forming protostars are still deeply embedded in very cold 
material, before eventually converting into the massive and warm cores now 
observed in the fully developed Orion nebula. 

\section{Implications for Star Formation}

We examine in this section the possibility of a triggered formation
for the protostellar cores through the onset of instabilities in the 
molecular gas surrounding the H{\small II} region. The efficiency 
of an instability depends on its characteristic growth rate, 
which obviously has to be compatible with the age of the nebula in order 
to perturb the overall stability of the gas. We first precise the dynamical 
age of the Trifid nebula before turning to the observational 
evidence for a triggered formation scenario and confront our observations 
with the predictions of a simple model. 

\subsection{The age of the nebula}

The size of the nebula suggests its expansion is only very recent. 
It has a radius 
of $2\pc$, approximately 5 times the radius of the Stromgren sphere created
by the exciting star (we assume a mean density of $2\times 10^3\cmmt$ in the
parental cloud, determined from the analysis of the large-scale CO
emission, Lefloch et al. 2000). From the
time-radius relation for evolved H{\small II} regions, we infer an
early dynamical age of 0.3-0.4~Myr for the Trifid. 
This agrees with the relatively high electron density 
measured~: 
150\cmmt\ (Lynds et al. 1985 ) versus 10-20\cmmt\ in the Rosette nebula 
(Celnik 1985). 
However, taking into account the duration of the
ultra-compact H{\small II} phase undergone by the exciting star HD164492A, 
more difficult to quantify but probably of the order of a few $10^5\yr$, 
the nebula is probably older and the expansion phase could
have lasted already long enough to leave ample time for condensations 
to collapse and form protostars.

\subsection{The shocked environment of TC3 and TC4}

From our observations, TC3 and TC4 are deeply embedded in a
layer of dense gas  and high-column density. When examining the 
morphology of the layer containing the TC4-TC4b-TC4c cores, we 
see that it is not much wider than the dust cores, with a typical 
length-to-width ratio of $\sim 7$ (Fig.~2). It is suggestive of a layer 
which would have fragmented into three condensations separated by 
$1.1-1.5\times 10^{17}\cm$. The high-density gas observations draw the
picture of a gas layer rather
uniform with a few condensations. 
The gas column densities  derived from CS and \ceio\ appear rather uniform 
across the gas~: 
the variations observed between the protostellar cores and the ambient 
gas not more than a factor of 2-3. For \htwo\ densities on the contrary, 
the differences are much more pronounced~: typically one order of magnitude. 
The emission of the dense gas detected appears to trace fragments of a 
larger shell surrounding the
ionized bubble (CL98). Moreover the layer is spatially anticorrelated with
the ambient gas, which is hardly detected in the high-density gas and column
density tracers. These observations provide evidence that the molecular layer
consists of some gas compressed and accelerated to a few $\kms$ in the
pre-cursor shock associated with the ionization front.

In the early stages of the expansion of the H{\small II} region, the
self-gravity of the molecular gas layer is negligible and the latter 
is confined by the external (thermal or ram) pressure
of the surrounding medium. As the layer builds up with some compressed 
material,  self-gravity starts to compete with the external pressure. 
This configuration is favorable to its fragmentation through the triggering
of ``Rayleigh-Taylor - like'' or ``Jeans-like'' instabilities. 
Accordingly, the TC3-TC4c may have formed from the  fragmentation of the 
compressed layer. 

In the molecular gas layers containing
TC3 and TC4 the linewidths are much larger than what is observed in 
the Taurus cloud for instance ($1.6\kms$ versus $0.5\kms$) where kinetic
temperatures of the same order can be found. 
Several authors have studied the stability of  a thin
plane-parallel gas layer confined by external pressure and self-gravity
(see e.g. Elmegreen \& Elmegreen 1978; Vishniac 1983; Elmegreen 1989;
Whitworth et al. 1994a,b). The results vary somewhat
depending on the assumed boundary conditions but show in all cases that
the first growing modes are purely hydrodynamical. 
In particular, MacLow \& Norman (1992) showed that the overstable modes 
which arise in the layer do not fragment it but generate some weak transverse 
shocks which can, in turn, generate some weak turbulence. 
It is therefore tempting to interpret 
these large non-thermal motions as the result of the  turbulence 
generated in the shocked gas as the shell expands.

\subsection{Was the birth of TC3 and TC4 triggered~?}

Previous work on the fragmentation of a compressed gas layer has shown the 
existence of two fast self-gravitating modes, a ``slow'' one on a timescale 
$\sim (G\rho_0)^{-1/2}$, where $\rho_0$ is the density of the ambient 
molecular cloud $\approx  2\times 10^3\cmmt$, 
and a ``fast'' one on a timescale $\sim (G\rho_0 M)^{-1/2}$ where M 
is the Mach 
number of the shocked layer relative to the effective sound-speed in the 
layer. In the case of the Trifid, the slowest mode 
has a typical timescale $\tau= 1.3Myr$, hence
cannot be efficient enough to trigger the condensation of the compressed layer
and allow the formation of protostars. 

As can be seen on Fig.~7, the velocity of the ambient
gas surrounding the compressed layer is $\sim 17-18\kms$, whereas the
velocity of the compressed layer containing TC4-4c is $\sim 22-23\kms$. 
Therefore, we adopt a typical expansion velocity of $5\kms$ and an effective 
sound speed $c_s$ of $0.6\kms$ in the shocked layer 
(the typical linewidth is $1.6\kms$ from our \ceio\ observations)~:   
we derive $M\simeq 8$ and a timescale of $4.6\times 10^5\yr$. This 
compares well to the age of the nebula, so that the fast 
instability could be  efficient enough to trigger the ``Jeans-collapse'' 
of the gas layer (Whitworth et al. 1994a,b). We now
estimate the physical parameters of the
fragments which would form.
The distance between two fragments is $L\sim c_s(G\rho_0 M)^{-1/2}= 9\times
10^{17}\cm$, and their mass is $\sim c_s^3(G^3 \rho_0 M)^{-1/2}= 24\msol$.    
The density in the shell is expected to be $M^2 \times \rho_0$ 
$\sim 64\times 2000=  10^5\cmmt$. 
 The typical fragment mass varies like $c_s^3$, hence even
a difference of a factor of 2-3 with the mass of TC4b and TC4c can be 
regarded as satisfactory. 

Although the derivation of all these figures is subject to the
uncertainties in the actual values of the parameters, the good 
agreement between the predicted properties of the fragments and 
the observational constraints from the properties of TC3-TC4c and the 
dynamical age of the Trifid is intriguing. 
If the instability studied here would have triggered the formation of 
TC3-TC4c, one would expect the velocity of the 
protostellar fragments to be close to that of the compressed layer, 
and differ somewhat from that of the unperturbed surrounding gas, possibly up 
to a few $\kms$. This is indeed observed towards TC3-TC4. 
As a conclusion, we cannot exclude that all the cores reported here might 
be the result of a spontaneous process rather than the fragmentation 
of the compressed layer in which they were found. There is no clear and 
direct evidence that the protostellar cores did form from the fragmentation 
of that layer. However, all the predictions about the physical properties 
of the fragments which can be made without a detailed
modeling of the Trifid (mass, size, velocity) fully agree with our 
observations. These are therefore good indications that 
such mechanism could be at work, and that the formation of the protostars 
would have been triggered by the expansion of the nebula. 

\subsection{Star formation on a large-scale}

The indications of star formation inside TC3 and TC4 suggest that
they are the cocoons of the most recent generation of stars
of the nebula. The physical conditions imply a typical 
Jeans mass $\sim 70\msol$, which is comparable to the mass of TC3 and TC4.
We speculate that the other cores TC4b-c 
and the subcondensations found in the envelope of TC3, smaller and less 
massive, are still in a pre-stellar stage, on the verge of collapsing.
We have indirect evidences that other stellar births took place in the
Trifid through the detection with ISOCAM of a hot
dust component of 2-3\arcsec\ in size which coincides spatially with
2 ultra-compact H{\small II} regions of B-type stars 
at only $6\arcsec$ from the exciting star of the Trifid (Lefloch et al. 2000). 

Although more work is needed to establish their nature, the point-like
sources detected close to the TC3 and TC4 cores in the $12\mu $m image are
probably some other newly born stars of the nebula.
TC3 and TC4 are the most recent events of a process which started with
the formation of the exciting star of the Trifid. Altogether these
observations suggest a continuous star forming process occuring on a scale
large enough to encompass the nebula, rather than a sequential process. 

Although compatible with a triggered formation scenario, our observations 
do not enable us to determine whether TC3 and TC4 were in a
quiescent stage before being hit by the shock wave, and subsequently started
to collapse, or if they were already on the path to form stars. 
However, they show in a rather unambiguous way massive protostellar cores
undergoing shock compression. High-angular observations are 
required to get more insight on this interaction and how severely the
collapse can be perturbed.

\section{Conclusions}

We have observed at millimeter and far-infrared wavelengths the molecular
condensation on the Southwestern border of the Trifid nebula, a young 
H{\small II} region.
Mapping of the dust thermal emission at 1.25mm reveals several massive 
condensations (between 8 and $90\msol$).
These condensations are made of very dense and cold gas and dust ($\rm T\sim
20\K$). Two of them, TC3 and TC4, exhibit unambiguous signs of star
formation activity and are associated with outflows. 

The dust and line properties of TC3 and TC4 are very similar to those of
some massive protostellar cores observed in Orion, once scaled at the 
distance of the H{\small II} region. 
The physical properties of their outflows are also similar to those powered by
intermediate and high-mass sources. The SiO abundances
we derive ($\sim 2\times 10^{-8}$) are also typical of outflows 
from young protostars and/or high-mass sources. 
TC4 might be associated with a point-like infrared source 
at 12$\mu$m; TC3 remains undetected at this wavelength and is a potential 
true high-mass protostar. 

The dust cores are embedded in a dense layer of molecular gas. The
physical properties of this component suggest it is made of material
compressed by the precursor shock of the ionization front driven by the
exciting star of the Trifid.
Based on a simple comparison of the large-scale properties of the dust cores
with analytical studies (mass and size), it seems plausible
that they formed from the fragmentation of the dense compressed layer.
Rather than TC3 and TC4 experiencing the shock ahead of the ionization front
while in the protostellar phase, the core formation and gravitational
collapse of these cores could have been triggered by the expansion of the 
nebula a few $10^5\yr$ ago. Our ISOCAM data suggest that the formation of 
TC3 and TC4 has been preceeded by other star formation episodes in the
whole nebula.

Several $10^5\yr$ later after the birth of the Trifid nebula, young massive 
protostars are still formed in the surrounding molecular gas. The 
protostellar cores of the Trifid provide us with a large wealth of information 
on the infancy  of now well-developed and well studied H{\small II} 
regions like Orion.

\begin{center} {\it Acknowledgements} \end{center}
We thank an anonymous referee for numerous comments in order to improve 
the manuscript. 
We acknowledge Spanish DGES for this research under grants PB96-0883 and 
ESP98-1351E.

\clearpage

\begin{table}
\caption[]{Physical properties of the protostellar dust cores 
compared with typical Class~0 sources. 
In the last column are given the source peak fluxes scaled 
to the distance of Taurus and $\rho$ Oph molecular clouds (160 pc).
}\begin{flushleft}
\begin{tabular}{lrrllcr}
\hline\noalign{\smallskip}
Name  & $\rm S_{1.25}^{peak}$ &
$\rm S_{1.25}^{int}{}^{(1)}$ &  $\rm M_t$ & N$_{\rm H}$ & $L_{bol}$ 
& $S^{*}$ \\
      & (Jy) & (Jy) &  ($\msol$) & ($\cmmd$) &  
($\lsol$)   & (Jy)  \\
\hline\noalign{\smallskip}

TC3           & 0.39  & 0.92 & 90       & 3.7(23) & \_     & 39  \\
TC4           & 0.29  & 0.60 & 58         & 2.8(23) & 520-2400    & 29  \\
TC4b          & 0.090 & 0.11 & 11         & 8.6(22) & \_     & 9    \\
TC4c          & 0.068 & 0.08 & 8         & 6.5(22) & \_     & 7   \\
              &       &      &            &         &        &      \\
VLA1623$^{1}$       & 0.77  & 1.5  & $\sim 0.9$ & 5-40(23)& 0.5-2.5& 0.8 \\
L1448-mm$^{2}$      & 0.58  & 0.58 & 2.0        & 3.4(23) & 9      & 2.0 \\
Cep\,E$^{3,4}$        & 0.46  & 1.11 & 18         & 7.9(23) & 75     & 9 \\
NGC7129$^{5}$       & 1.49  & 1.49 & 6          & 1.6(23) & 430    & 58  \\
\noalign{\smallskip}
\hline\noalign{\smallskip}   
\label{core}
\end{tabular}
\end{flushleft}

$^{(1)}$~Determined from the dust emission contour at $50\%$.\\
REFERENCES~: (1)~Andr\'e et al. 1993: (2)~Bachiller et al. 1991;
(3)~Lefloch et al. 1996: (4)~Noriega-Crespo et al. 1998;
(5)~Eiroa et al. 1998.
\end{table}

\begin{table*}
\caption[]{Physical properties of TC3 compared with a few Orion protostellar
  cores. }
\begin{flushleft}
\begin{tabular}{lcccc}
\hline\noalign{\smallskip}%
                            &  TC3      &  IRc2$^{(0)}$ & CS1$^{(0)}$   & CS3$^{(0)}$    \\
Continuum Source Size (pc)  & 0.22 & 0.02 & 0.02  & 0.02    \\
CS Source Size  (pc)        & 0.14 & 0.02 & 0.02 & 0.05    \\
Dust Core Mass ($\msol$)    & 90 & 125 &  45 & 54  \\
Dust temperature ($\K$)     & 20  & 100 &  70 & 60       \\
CS $\rm T_b^{21}$ peak scaled at $d=500\pc$      & 20 & 57  & 22  & 16 \\
SiO $\rm T_b^{21}$ peak scaled at $d=500\pc$     & 9  & 17  & \_ & 5 \\
CS Core Mass ($\msol$)      & 12  & 80 & 6 & 12 \\
CS Core Density ($\cmmt$)      & $1-2\times10^6$ & $3-6\times 10^7$ &
$2-13\times 10^6$ & $0.5-5\times 10^6$        \\
\hline\noalign{\smallskip}                     
\end{tabular}  
\end{flushleft}
$^{0}$~data taken from Mundy et al. (1986) and Ziurys et al. (1990)\\
\label{compar}
\end{table*}

\begin{table*}
\caption[]{Physical properties of the TC3 and TC4 molecular outflows, as 
derived from the SiO(2-1) line under the conditions of LTE and assuming an
excitation temperature of 60~K. 
}
\begin{flushleft}
\begin{tabular}{lllll}
\hline\noalign{\smallskip}%
\hline\noalign{\smallskip}\\
Parameters                     & TC3    & TC4 \\
\hline\noalign{\smallskip}\\
Outflow mean density ($\cmmt$) & $2.5-5\times 10^5$  & $3-5\times 10^{5}$ \\
Mass  $M$   ($\msol$)          & 1.1    & 1.5\\
Linear Momentum $P$ ($\msol\kms$) & 10  & 11 \\
Dynamical age $\tau_d$ ($\yr$)    & $6.8\times 10^3$   & $6.8\times 10^3$ \\
Mass-Loss Rate $\dot{M}= M/\tau_d$ ($\msol\yrmu$) & 
$1.7\times10^{-4}$ & $2.2\times 10^{-4}$ \\
Mechanical Energy $E$ (erg)            & $1.0\times 10^{45}$ & $9 \times 10^{44}$\\
Mechanical Luminosity $L= E/\tau_d$  ($\lsol$)  & 1.2       & 1.1\\
\hline\noalign{\smallskip}
\end{tabular}
\end{flushleft}
\label{outflow}
\end{table*}

{\small\noindent Fig. 1 --
$\rm H\alpha$ image of the Trifid nebula taken at the IAC80 telescope
(see CL98). We have superposed the  
1250$\mum$ continuum emission observed with the IRAM 30m-telescope, which 
coincides with the dust lanes of the nebula and the surrounding molecular 
shell. The region studied in the article is marked with the box. }

{\small\noindent Fig. 2 --
Map of the 1250$\mum$ continuum emission in the
Southwestern part of the Trifid nebula observed with the IRAM
30m-telescope, superposed upon the $\rm H\alpha$ emission of the region
(greyscale) . 
The millimeter emission has been convolved with a gaussian $11\arcsec$ HPFW
and degraded to a resolution of $15\arcsec$. Contour levels are
5, 10, 15, 20, 30, 45, 60 to 150 by 30, 200 to 350 by 50 mJy/beam.
The units of the axis are in arcsec offsets with
respect to the position of exciting star of the nebula~: 
$\alpha_{2000}= 18^{\rm h} 02^{\rm m} 23.55^{\rm s}$
 $\delta_{2000}= -23^{\circ} 01\arcmin 51\arcsec$.
The position of the different cores is indicated by arrows. 
}

{\small\noindent Fig. 3 --
Continuum emission (greyscale) observed at $12\mum$ with ISOCAM towards 
TC3-TC4. The original map was centered on the exciting star of the Trifid 
($\alpha_{2000}= 18^{\rm h} 02^{\rm m} 23.55^{\rm s}$
 $\delta_{2000}= -23^{\circ} 01\arcmin 51\arcsec$). The axis are in 
arcsec offsets with respect to this position. 
The cores are located in the region of minimum emission. There is 
one possible IR point-like source associated with TC4.
We show in contours the $1250\mu m$ flux distribution observed with 
the IRAM 30m-telescope.
}

{\small\noindent Fig. 4 --
{\em top})~Far-infrared emission in the range $45-197\mu m$ observed 
with the ISO LWS spectrometer towards TC4 (thick) and a reference 
position outside the molecular gas region (thin), at 
$\alpha= 18^{h}02^{m}35.1^{s}$ $\delta= -23^{\circ}03'52''$ (J2000). 
The spectrum is dominated by ionic and atomic lines excited in the PDR 
and the ionized nebula. We show (thin line) the fit of the emission 
towards TC4 from two modified black-bodies at 18~K and 40~K (see text). 
{\em bottom})
Spectral Energy Distribution of TC4 after subtracting the contribution of
the PDR, observed throughout the whole nebula. We have
indicated (square) the $1250\mum$ flux (integrated over one ISO beam) 
and the  $12\mu m$ flux of the IR source near the TC4 millimeter dust peak.
The emission of the extended dust component was estimated from the
reference position and subtracted to the LWS spectrum.
} 
\label{fig4}
\smallskip

{\small\noindent Fig. 5 --
Montage of spectra in the different millimeter lines observed towards the
TC3 and TC4 dust cores (left and right panel respectively). The 
intensities are in main-beam brightness
temperatures. The wings of the outflows (nicely defined in the SiO
\jthreetotwo\ line), are very broad and range from 
-20 to $+60\kms$.
}
\label{fig5}
\smallskip

{\small\noindent Fig. 6 --
Map of the main-beam brightness temperature in the CS \jthreetotwo\ line.
First contour and contour interval are 0.25~K. The contour at half-power 
is marked in thick.}
\label{fig6}
\smallskip

{\small\noindent Fig.~7 --
Channel map of the $\ceio\, \jonetozero$ velocity-integrated emission, 
between 17.5 and $23\kms$. Contours are 0.25, 0.5, 1, 1.5, etc... up to 
$5\K\kms$.  The position of the exciting star HD~164492A and of 
protostars TC3 and TC4 is indicated by a white
mark. 
}
\label{fig7}
\smallskip

{\small\noindent Fig. 8 --
Velocity-integrated emission map of the SiO \jtwotoone\ line (contours) 
superposed on the 1.25mm continuum emission (greyscale).  
The emission was integrated between -10 and $18.5\kms$ for the blue
wing (solid contours), and from 24 to $60\kms$ for the red wing (dashed 
contours). Contours range from 25\% to 95\% of the maximum peak in each wing.
}
\label{fig8}
\smallskip

\end{document}